\mag=\magstep1
\documentclass[twoside,reqno]{amsart}
\pdfoutput=1
\usepackage{color}
\usepackage[top=3.8cm, bottom=4cm, left=3.8cm, right=3.8cm]{geometry}
\usepackage{hyperref}
\usepackage{fancyhdr}
\usepackage{fmtcount}
\usepackage[norelsize,lined,boxed,linesnumbered,algosection]{algorithm2e}
\usepackage{graphicx}
\usepackage[latin1]{inputenc}
\usepackage{amssymb}
\usepackage{tikz}
\usetikzlibrary{matrix}
\usepackage{amsthm}
\usepackage{amsmath}
\usepackage{appendix}
\usepackage{amsfonts}
\usepackage{array}
\usepackage{enumerate}
\usepackage{csvsimple}
\usepackage{booktabs}
\usepackage{longtable}
\usepackage{float}
\usepackage{enumerate}
\usepackage{mathrsfs}
\usepackage{footnote}
\usepackage{tablefootnote}
\usepackage{graphicx}
\usepackage{multirow}
\usepackage{subfig}
\usepackage[all]{hypcap}
\usepackage[font=it]{caption}
\usepackage{myMacro}
\usepackage[protrusion=true,expansion=true]{microtype} 
\SetAlCapSkip{0.5em}

\SetKwInOut{KwInput}{Input}
\SetKwInOut{KwOutput}{Output}

\hypersetup{pdftitle=Crunching Mortality and Life Insurance Portfolios with Extended CreditRisk$^+$, pdfauthor=Jonas Hirz,pdfdisplaydoctitle=true,pdffitwindow=true}

\numberwithin{equation}{section}
\numberwithin{table}{section}
\numberwithin{figure}{section}

\copyrightinfo{}{The authors}

\begin{document}

\pdfpageheight=297 true mm
\pdfpagewidth=210 true mm
\pdfhorigin=1 true mm
\pdfvorigin=4 true mm

\title[Crunching Mortality and Life Insurance Portfolios]
{Crunching Mortality and Life Insurance Portfolios with Extended CreditRisk$^+$}


\thanks{J.~Hirz gratefully acknowledges financial support from the Australian Government via the 2014 Endeavour Research Fellowship, as well as from the Oesterreichische Nationalbank (Anniversary Fund, project number: 14977) and Arithmetica. 
P.~V.~Shevchenko gratefully acknowledges financial support by the CSIRO-Monash
Superannuation Research Cluster, a collaboration among CSIRO, Monash University, Griffith University, the University of Western Australia, the University of Warwick, and stakeholders of the retirement system in the interest of better outcomes for all. Furthermore, we thank Anselm Fleischmann (BELTIOS GmbH, Vienna) for fruitful discussions.}

\author[J.~Hirz]{Jonas~Hirz}
\address[Jonas~Hirz]{Department of Financial and Actuarial Mathematics, TU Wien, Wiedner Hauptstr. 8/105-1, 1040 Vienna, Austria}
\email{hirz@fam.tuwien.ac.at}

\author[U.~Schmock]{Uwe~Schmock}
\address[Uwe~Schmock]{Department of Financial and Actuarial Mathematics, TU Wien, Wiedner Hauptstrasse 8-10, 1040 Vienna, Austria}
\email{schmock@fam.tuwien.ac.at}

\urladdr{\href{http://www.fam.tuwien.ac.at/~schmock/}
{http://www.fam.tuwien.ac.at/\~{}schmock/}}

\author[P.~V.~Shevchenko]{Pavel~V.~Shevchenko}
\address[Pavel~V.~Shevchenko]{Applied Finance and Actuarial Studies, Macquarie University, NSW, Australia}
\email{pavel.shevchenko@mq.edu.au}


\date{\today}

\renewcommand{\subjclassname}{\textup{2\,000} Mathematics Subject
     Classification}

\begin{abstract}
	 This paper provides a summary of actuarial applications in life insurance 
	of the collective risk model  extended CreditRisk$^+$ amongst which we find stochastic
mortality modelling, joint modelling of underlying death causes as well as profit and loss modelling of life insurance and annuity portfolios
as required in (partial) internal models. This approach provides an efficient, numerically stable
algorithm for an exact calculation of the one-period profit and loss distribution where various
sources of risk can be considered. Model parameters can be estimated by likelihood optimization and Bayesian approaches, using
Markov-Chain Monte-Carlo techniques. We provide real world examples using
Austrian and Australian death data.
\end{abstract}

\maketitle

\section{Introduction}

Pricing of retirement  income products, using internal models and calculating P\&L attributions in corresponding portfolios depend crucially on the accuracy of predicted death probabilities.
Life insurers and pension funds typically use deterministic generation life tables obtained
from crude death rates and then apply some forecasting model or generation life tables. Afterwards, artificial risk margins are often added to account for phenomena associated with longevity, size of the portfolio, selection phenomena, estimation and various other sources.
	These approaches often lack a stochastic foundation and are certainly not consistently appropriate
	for all companies due to a possibly twisted mix of these sources of risk.
	\begin{figure}[htp]
		\begin{center}
		\captionsetup{aboveskip=0.2\normalbaselineskip}
		\includegraphics[width=0.9\textwidth]{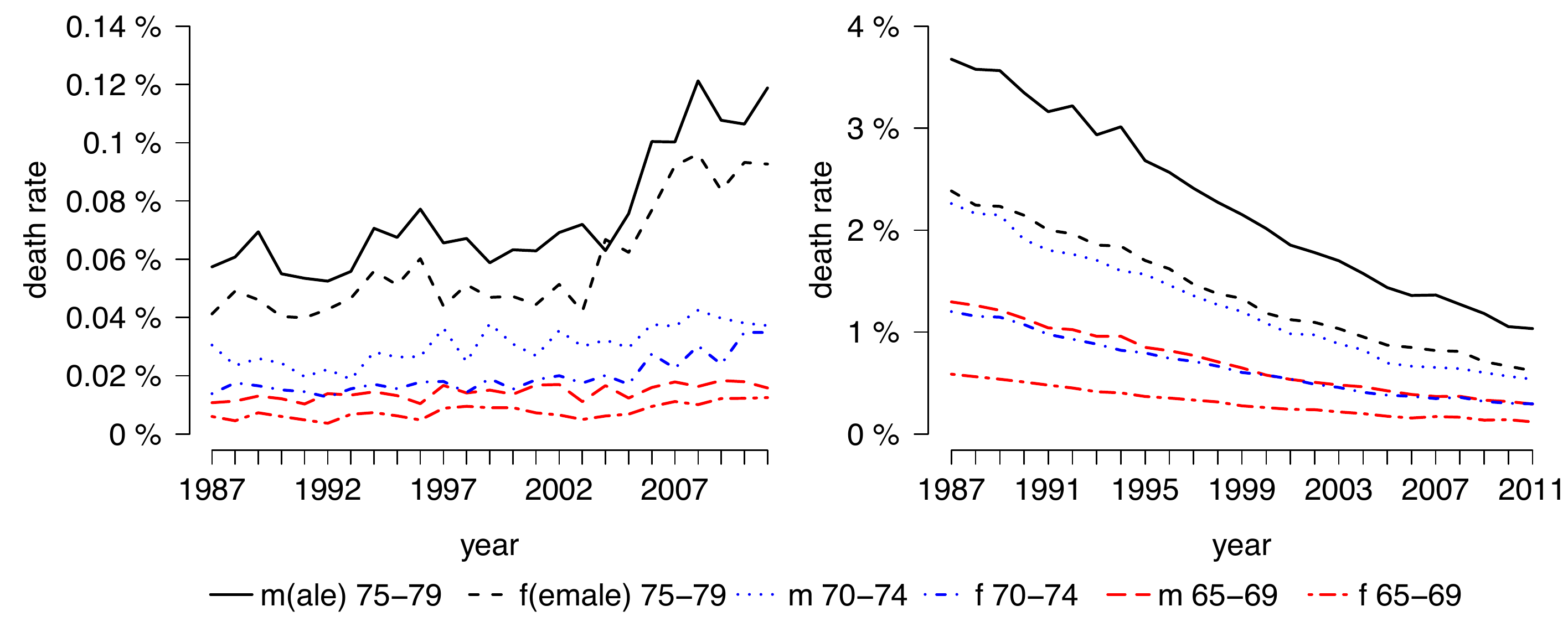}
		\caption[Australian death rates for mental and behavioural disorders 1987--2011]{Australian death rates for mental and behavioural disorders (left) and for 
		circulatory diseases (right).}
		\label{fig:death_causes}
		 \end{center}
	\end{figure}
	Moreover, we have observed drastic shifts in death rates (yearly deaths divided by population at June 30~in the corresponding year) due to certain underlying  death causes over the past decades. An underlying death cause is to be understood as
	the disease or injury that initiated the train of morbid events leading directly to
	death. As an 
	illustration of this fact, Figure \ref{fig:death_causes} shows death rates based on Australian data for death causes, such as mental and behavioural disorders and
 circulatory diseases, from 1987 to 2011 for various 
	age categories and both genders. Diseases of the circulatory system, such as ischaemic heart disease, have been clearly 
	reduced throughout the past years while death rates due to mental and behavioural disorders, such as dementia, have doubled for older age groups. 
	This observation nicely illustrates the existence of serial dependence amongst different 
	death causes.
	
	In our recent work Hirz, Schmock and Shevchenko \cite{ourPaper}, we are
	introducing a novel actuarial framework using a collective risk model called extended
CreditRisk$^+$. In light of the issues described above, it provides a unified and stochastically sound approach to model mortality on the one hand and risk aggregation in annuity as well as life insurance portfolios on the other hand. The general form of this credit risk model was introduced by Schmock
\cite{schmock} and is very different from usual time series approaches such as
Lee--Carter or cohort models, see Lee and Carter \cite{Lee-Carter}, as well as Cairns et al.~\cite{cairns2009}. Stochastic modelling of mortality has become increasingly important also because new regulatory requirements such as
Solvency II allow internal stochastic models for a more risk-sensitive evaluation of 
capital requirements and the ability to derive accurate P\&L attributions.
 Our model allows multiple applications, including stochastic modelling of mortality, joint forecasting of death cause intensities, calculation of P\&L in life insurance portfolios exactly via Panjer recursion instead of a Monte Carlo method, and partial internal model applications in the (biometric) underwriting risk module. In particular, our model is able to quantify the risk of statistical fluctuations within the next period (experience variance) and parameter uncertainty (change in assumptions). 

In this paper we focus on concepts and application aspects of the proposed framework while for details of the proofs and precise mathematical conditions we refer to Hirz, Schmock and Shevchenko \cite{ourPaper}. We start with a brief model description in Section \ref{sec:model} and 
recall different estimation procedures in Section \ref{sec:estimation}. Then, 
in Section \ref{sec:real}, we apply our model and estimation procedures to Australian and Austrian
data and give some further applications including mortality forecasts, scenario analysis and a potential internal model application. In Section \ref{validation} we give some validation techniques and end with  concluding remarks in Section \ref{sec:conclusion}. 
	
	\section{Model}\label{sec:model}
	
Based on the collective risk model \emph{extended CreditRisk$^+$}, see Schmock \cite{schmock}, we introduce our model as follows.
	Let $\{1,\dots,m\}$ denote
	the set of people (we call them \emph{policyholders} in light of potential applications to annuity portfolios)  in the portfolio
	and let random death indicators $N_1,\dots,N_m$
	indicate the \emph{number of deaths} (this may also include or be lapse) of each policyholder in the following period.
	The event $\{N_i=0\}$ indicates survival of person $i$ whilst $\{N_i\geq 1\}$ indicates death. Thus, we define death probability $q_i:=\expv{N_i}$.
		In reality, death indicators are Bernoulli random variables.
	Unfortunately in practice, such an approach is
	not tractable for calculating loss distributions of large portfolios as execution times
	explode.
	Instead, we will assume a mixed Poisson distribution for each policyholder which is very accurate (as can be shown in numerical examples) and which provides an efficient way for calculating loss distributions.
	
	Consider
	independent \emph{portfolio quantities} ${Y}_1,\dots, Y_m$ with $d\geq 1$ dimensions which are assumed to be independent of ${N}_1,\dots,{N}_m$. We are then interested in (random) cumulative quantities due to deaths
	\[
      		S:=\sum_{i=1}^{m}\sum_{j=1}^{N_{i}}\boldsymbol{Y}_{i,j}\,,
	\]
	where $({Y}_{i,j})_{j\in\na}$ for every $i$ is an i.i.d.~sequence of random variables
		with the same distribution as ${Y}_{i}$.
	Thus, depending on the specific application of our proposed model, portfolio quantities and the sum $S$ can be interpreted differently:
	\begin{enumerate}[(a)]
		\item For applications in the context of internal models we may set $Y_i$ as the best estimate 
	liability (BEL) of the contract of $i$ such that $S$ gives the sum of BELs of all policyholders who do not survive the following period.
	\item In the context of portfolio payment analysis we may set $Y_i$ as the payments (such as annuities)
		to $i$ over the next period such that $S$ gives the sum of payments to all policyholders who do not survive the following period. We may include premiums in a second dimension in order 
		to get joint distributions of premiums and payments.
		\item For for applications in the context of mortality and population estimation we may set $Y_i=1$ such that $S$ gives the number of deaths in the next period.
	\end{enumerate}
	In the context of cumulative payments $L$ in an annuity portfolio, for example, it is necessary to consider 
		the sum the payments if everyone survives minus all payments which need not be paid due to deaths, i.e.~$L:=\sum_{i=1}^{m}\boldsymbol{X}_{i}-\boldsymbol{S}$, where $X_i$ may be equal to $Y_i$ or some (random) fraction of sub-annual payments are considered.
		
		To ensure
	multi-level dependence, we introduce stochastic risk factors $\Lambda_{1},\dots,\Lambda_{K}$ which are independent
		         	and gamma distributed with mean one and
				variance $\sigma_k^2> 0$. 
				These risk factors can be seen as latent random factors which jointly influence deaths of all policyholders, such as underlying death causes with the stochasticity lying in changes in treatments, better medication, or outbreaks of epidemics, for example. Both the assumption of independence amongst risk factors and the assumption of gamma distributions can be relaxed, see Schmock \cite{schmock}.
				Constants $w_{i,0},\dots,w_{i,K}\in[\4 0,1]$ for policyholder $i$ represent weights
				with
	$w_{i,0}+\dots +w_{i,K}=1$, indicating the vulnerability of policyholder $i$ to risk factors. For every policyholder $i$,
			the total number of deaths $N_i$ is split up according
			to risk factors as
			$N_i=N_{i,0}+\dots+N_{i,K}$,
			where  the idiosyncratic $N_{1,0},\dots,N_{m,0}$ are independent
			from one another, as well as all other
				random variables, and Poisson distributed with
				intensity $q_i\4 w_{i,0}$. Furthermore, conditional on the  risk factors, death indicators $\{N_{i,k}\}_{i\in\{1,\dots,m\},k\in\{1,\dots,K\}}$ are  independent and Poisson distributed with random intensity $q_{i}\4 w_{i,k}\4 \Lambda_{k}$.
	Common stochastic risk factors introduce dependence such that
	$\cov(N_i,N_j)= q_i\4 q_j\4\sum_{k=1}^K w_{i,k}\4 w_{j,k}\4 \sigma_k^2 $, for all $i\neq j$.

There exists a numerically stable algorithm to derive the distribution 
of ${S}$ very efficiently. It can be found in the detailed paper of Hirz, Schmock and Shevchenko \cite{ourPaper}, or in the lecture notes of Schmock \cite[Section 6.7]{schmock}. 
Basically, this algorithm uses iterated Panjer recursion to derive\/ $\Prob(\boldsymbol{S}=n)$ up to every desired $n\in\na$. 
Approximations arise from mixed Poisson instead of Bernoulli distributions for deaths and---if required---due to stochastic rounding to work with greater loss units. Nevertheless, implementations of 
			this algorithm are significantly faster than Monte Carlo approximations for comparable error levels. 
						\begin{table}[ht]
	\begin{center}\footnotesize{
		\caption{Quantiles, execution times (speed) and total variation distance (accuracy) of Monte Carlo with $50\,000$ simulations, as well as our annuity model given a simple portfolio.} 
		\label{table_est}
		\begin{tabular}{r|rrrrr|rr}
		\cline{2-6}\rule[-6pt]{0pt}{18pt}
			& \multicolumn{5}{c|}{quantiles}  & &  \\ 
			\cline{2-8}\rule[-6pt]{0pt}{18pt}
			& 1\%  & 10\% & 50\%& 90\%  & 99\%  & speed & accuracy \\ 
			\hline\rule[10pt]{0pt}{0pt}
			Monte Carlo, $w_{i,0}=1$ & $450$ & $472$ & $500$ & $528$ &  $552$ & $22.99$ sec. & $0.0187$\\\rule[-5pt]{0pt}{0pt}
			annuity model, $w_{i,0}=1$ & $449$ & $471$ & $500$ & $529$ & $553$ & $0.01$ sec. & $0.0125$\\
			\hline\rule[10pt]{0pt}{0pt}
			Monte Carlo, $w_{i,1}=1$ & $202$ & $310$ & $483$ & $711$ & $936$ & $23.07$ sec. & $0.0489$\\\rule[-5pt]{0pt}{0pt}
			annuity model, $w_{i,1}=1$ & $204$ & $309$ & $483$ & $712$ & $944$ & $0.02$ sec. &$ \leq 0.0500$\\\hline
		\end{tabular}}
	\end{center}
\end{table}
			As an illustration we take a portfolio with $m=10\,000$ policyholders having death  probabilities $q:=q_i=0.05$ and payments $Y_i=1$. 
			We then derive the distribution of $S$ using our annuity model for the case with just idiosyncratic risk, i.e., $w_{i,0}=1$, and for the case with just one common stochastic risk factor $\Lambda_1$ with variance $\sigma_1=0.1$ and no idiosyncratic risk, i.e., $w_{i,1}=1$. 
			Then, using $50\, 000$ simulations of the corresponding model where $N_i$ is 
			Bernoulli distributed or mixed Bernoulli distributed given truncated risk factor $\Lambda_1\conditioned \Lambda_1\leq \frac{1}{q}$, we compare the results of our annuity model to Monte Carlo, respectively. Truncation of risk factors in the Bernoulli model is necessary as otherwise death probabilities may exceed one. 
			We observe that our annuity model drastically reduces execution times at comparable error levels. Error levels 
			in the purely idiosyncratic case are 
			measured in terms of total variation distance between approximations and 
			the binomial distribution with parameters $(10\,000,0.05)$ which arises as 
			the independent sum of all Bernoulli random variables. 
			Error levels in the purely non-idiosyncratic case are 
			measured in terms of total variation distance between approximations and 
			the mixed binomial distribution where for our annuity model we use Poisson approximation to get an upper bound. Results are summarised in Table \ref{table_est}.			

	To account for trends in death probabilities and shifts in death causes, we introduce
	the following parameter families. 
	First, let $F^{\mathrm{Lap}}$ denote the cumulative Laplace distribution 
	function with mean zero and variance two, i.e.,
	\begin{equation}\label{LaplaceDistr}
		F^{\mathrm{Lap}}(x)=\frac{1}{2}+\frac{1}{2}\sign(x)\4\big(1-\exp(-|x|)\big)\,,\quad x\in\re\,,
	\end{equation}
	and trend acceleration and trend reduction with parameters $(\zeta,\eta)\in\re\times(0,\infty)$ is given by $\mathcal{T}_{\zeta,\eta}(t)=(\mathcal{T}^*_{\zeta,\eta}(t)-\mathcal{T}^*_{\zeta,\eta}(t_0))/(\mathcal{T}^*_{\zeta,\eta}(t_0)-\mathcal{T}^*_{\zeta,\eta}(t_0-1))$ with normalisation 
	parameter $t_0\in\re$ and 
	\begin{equation}\label{CauchyDistr}
		\mathcal{T}_{\zeta,\eta}(t)=\frac{1}{\eta}\arctan(\eta\4(t-\zeta))\,,\quad t\in\re\,.
	\end{equation}
	The normalisation above guarantees that parameter $\beta$ can be compared across 
	different data sets. 
	Given $x< 0$, (\ref{LaplaceDistr}) simplifies to an exponential function. 
	Then, death probabilities for all policyholders $i$, with year of birth $z_i\in\re$, is
	given by
	\begin{equation}\label{eq:PDFamily}
		q_{i}(t)=F^{\mathrm{Lap}}\big(\alpha_{i} +\beta_{i}\4 \mathcal{T}_{\zeta_{i},\eta_{i}}(t)+\kappa_{z_i}\big)\,,
	\end{equation}
	where\/ $\alpha_{i},\beta_{i},\zeta_{i},\kappa_{z_i}\in\re$
	and\/ $\eta_{i}\in(0,\infty)$, as well as, for\/ $k\in\{0,\dots,K\}$, weights 
	are given by
	\begin{equation}\label{eq:WeightFamily}
	w_{i,k}(t)= \frac{\exp\big(u_{i,k}+v_{i,k}\4 \mathcal{T}_{\phi_{k},\psi_{k}}(t)\big)}{\sum_{j=0}^K \exp\big(u_{i,j}+v_{i,j}\4\mathcal{T}_{\phi_j,\psi_j}(t)\big)}\,,
	\end{equation}
	with\/ $u_{i,0},v_{i,0},\phi_0,\dots,u_{i,K},v_{i,K},\phi_K\in\re$, as well as\/ $\psi_{0},\dots,\psi_{K}\in(0,\infty)$. 
	Thus, 
	in practical situations, this yields 
	an exponential evolution of death probabilities in (\ref{eq:PDFamily}) modulo trend reduction 
	and cohort effects.
	Expression (\ref{CauchyDistr}) is used for a trend reduction technique which is motivated by Kainhofer, Predota and Schmock \cite[Section 4.6.2]{AVOE_annuity_table} and ensures that weights and death probabilities are limited 
	as $t\to\infty$.
		Conceptionally, the parameter $\eta$ gives the speed of trend reduction and the parameter 
		$\zeta$ on the other hand gives the shift on the S-shaped arctangent curve, i.e., the 
		location of trend acceleration and trend reduction.
		Parameter $\kappa$ models cohort effects for groups with the same or a similar year of birth. 
		This factor can also be understood in a more general context, in the sense that a 
		cohort effect may model categorical variates such as smoker/non-smoker, diabetic/non-diabetic 
		or country of residence. Moreover, cohort effects could be used for modelling weights $w_{i,k}(t)$ but is avoided here as sparse data does not allow proper estimation. Even more restrictive, 
		in applications we often fix $\phi$ and $\psi$ to reduce dimensionality to suitable levels.
		Note that in our model every other deterministic trend family can be assumed.

		\section{Estimation}\label{sec:estimation}
		Considering discrete-time periods\/ 
	$U:=\{1,\dots,T\}$ with time index $t\in U$, we assume that age- and calendar year-dependent death probabilities\/ $q_i(t)$ and corresponding weights $w_{i,k}(t)$, see (\ref{eq:PDFamily}) and (\ref{eq:WeightFamily}), are the same for all 
			representative policyholders\/ $i\in\{1,\dots,m\}$ within 
			the same age category\/ $a\in\{1,\dots, A\}$, same gender\/ $g\in\{\mathrm{f},\mathrm{m}\}$ and with respect to the same risk factor\/ 
			$\Lambda_k(t)$ with death cause\/ $k\in\{0,\dots,K\}$. For notational purposes we may therefore define\/ 
			$q_{a,g}(t):=q_i(t)$ and\/ 
			$w_{a,g,k}(t):=w_{i,k}(t)$ for a representative policyholder\/ $i$ of age category\/ $a$ and gender\/ $g$ with respect to 
			risk factor\/ $\Lambda_k(t)$, with birth years being denoted by $z_{a,t}$. All random variables at time\/ $t\in\{1,\dots,T\}$ 
			are assumed to be independent of random variables at some different point in time\/ $s\neq t$ with\/ $s\in\{1,\dots,T\}$, as well as  risk factors\/ $\Lambda_k(1),\dots,\Lambda_k(T)$ are identically distributed for each $k$.
			Given historical population counts\/ $m_{a,g}(t)$ and
	historical number of deaths\/ $n_{a,g,k}(t)$
due to underlying death causes\/ $k=0,\dots,K$ we can then derive various estimation procedures if
	$n_{a,g,k}(t)$ is assumed to be a realisation of the random variable
	\[
		N_{a,g,k}(t):=\sum_{i\in M_{a,g}(t)}N_{i,k}(t)\,,
	\]
	where\/ $M_{a,g}(t)\subset\{1,\dots,m(t)\}$ with\/ $|M_{a,g}(t)|=m_{a,g}(t)$ denotes the set of   policyholders 
	of  specified age group and gender. 
	Then, the likelihood function\/ $\ell( \boldsymbol{n}\conditioned\theta_q,\theta_w,\boldsymbol{\sigma})$
	of parameters\/ $\theta_q:=({\alpha},{\beta},\zeta,\eta,\kappa)$, as well as\/
	$\theta_w:=({u},{v},\phi,\psi)$ and\/ ${\sigma}:=(\sigma_k)$
	 given data\/
	$ \boldsymbol{n}:=(n_{a,g,k}(t))$
	is given by
	\begin{equation}\label{MLE_likelihood}
	\begin{split}
		\ell&(n\conditioned\theta_q,\theta_w,\boldsymbol{\sigma})=
		\prod_{t=1}^T \Bigg(\bigg(\prod_{a=1}^A\prod_{g\in\{\mathrm{f},\mathrm{m}\}}\frac{e^{-\rho_{a,g,0}(t)}\4 \rho_{a,g,0}(t)^{n_{a,g,0}(t)}}
		{n_{a,g,k}(t)!}\bigg)\\
		&\times		
		\prod_{k=1}^K\bigg(\frac{\Gamma(1/\sigma^2_k+n_k(t))}
		{\Gamma(1/\sigma^2_k)\4 (\sigma^2_k)^{1/\sigma^2_k}\4 (1/\sigma^2_k+\rho_k(t))^{1/\sigma^2_k+n_k(t)}}\prod_{a=1}^A\prod_{g\in\{\mathrm{f},\mathrm{m}\}}
		\frac{ \rho_{a,g,k}(t)^{n_{a,g,k}(t)}}
		{ n_{a,g,k}(t)!}\bigg)\Bigg)\,.
	\end{split}
	\end{equation}
	where	$n_k(t):=\sum_{a=1}^A\sum_{g\in\{\mathrm{f},\mathrm{m}\}}n_{a,g,k}(t)$,
	as well as\/ $\rho_{a,g,k}(t):=m_{a,g}(t)\4 q_{a,g}(t)\4 w_{a,g,k}(t)$  and
	$\rho_k(t):=\sum_{a=1}^A\sum_{g\in\{\mathrm{f},\mathrm{m}\}}\rho_{a,g,k}(t)$.
	
	Since the products in (\ref{MLE_likelihood}) can become small, we recommend to use the log-likelihood function instead. Examples suggest that maximum-likelihood estimates are unique. However, deterministic numerical optimisation routines easily break down due to high dimensionality. Switching to a Bayesian setting, it is straightforward to apply Markov chain Monte Carlo (MCMC) methods to get many samples
$\boldsymbol{\theta}=(\theta^i_q,\theta^i_w,\boldsymbol{\sigma}^i)$ from the posterior distribution
$\pi(\boldsymbol{\theta}|\boldsymbol{n})\propto \ell( \boldsymbol{n}|\boldsymbol{\theta})\pi(\boldsymbol{\theta})$ where $\pi(\boldsymbol{\theta})$ denotes the prior distribution of parameters.
The mean over these samples provides a good point estimate for model parameters and sampled posterior distribution can be used to estimate parameter uncertainty.
Various MCMC algorithms are available. Hirz, Schmock and Shevchenko \cite{ourPaper} implement a well-known random walk Metropolis--Hastings within Gibbs algorithm, in which case the mode of the posterior samples corresponds to a maximum-likelihood estimate. The method requires a certain burn-in period until the generated chain becomes stationary. To reduce long computational times, one can run several independent MCMC chains with different starting points on different CPUs in a parallel way.
To prevent overfitting, it is possible to regularise, i.e., smooth, maximum a posteriori estimates via adjusting the prior distribution $\pi(\boldsymbol{\theta})$. This technique is particularly used in regression, as well as in many applications, such 
		as signal processing. When forecasting death probabilities in Section \ref{sec:forecaDP}, we use a Gaussian prior distribution with a certain correlation structure.
		
	If risk factors are not integrated out, under a Bayesian setting, we may also derive 
	the posterior distribution of the risk factors. Necessarily in that case, 
	estimates\/  
	$\boldsymbol{\hat{\lambda}}$ for risk factors and\/
	$\boldsymbol{\hat{\sigma}}$ for risk factor variances
	satisfy
	\begin{equation}\label{lambda_MAP}
		\hat{\lambda}^{\mathrm{MAP}}_k(t)=\frac{1/(\hat{\sigma}_k^{\mathrm{MAP}})^2-1+\sum_{a=1}^A\sum_{g\in\{\mathrm{f},\mathrm{m}\}}n_{a,g,k}(t)}
	{1/(\hat{\sigma}_k^{\mathrm{MAP}})^2+\sum_{a=1}^A\sum_{g\in\{\mathrm{f},\mathrm{m}\}}\4 \rho_{a,g,k}(t)}
	\end{equation}
	if\/ $1/(\hat{\sigma}_k^{\mathrm{MAP}})^2-1+\sum_{a=1}^A\sum_{g\in\{\mathrm{f},\mathrm{m}\}}n_{a,g,k}(t)> 0$, as well as
	\begin{equation}\label{beta_MAP}
		2\4\log \hat{\sigma}_k^{\mathrm{MAP}}+\frac{\Gamma'\big(1/(\hat{\sigma}_k^{\mathrm{MAP}})^2\big)}{\Gamma\big(1/(\hat{\sigma}_k^{\mathrm{MAP}})^2\big)}
		=\frac{1}{T}\sum_{t=1}^T\big(1+\log\hat{\lambda}^{\mathrm{MAP}}_k(t)-\hat{\lambda}^{\mathrm{MAP}}_k(t)\big)\,,
	\end{equation}
	where, for given\/ $\hat{\lambda}^{\mathrm{MAP}}_k(1),\dots,\hat{\lambda}^{\mathrm{MAP}}_k(T)>0$, (\ref{beta_MAP}) 
	has a unique solution which is strictly positive. 
	Thus, rougher---but still very accurate---estimates are given by
	\begin{equation}\label{MAPappr_lambda}
		\hat{\lambda}^{\mathrm{MAPappr}}_k(t):=\frac{-1+\sum_{a=1}^A\sum_{g\in\{\mathrm{f},\mathrm{m}\}}n_{a,g,k}(t)}
		{\sum_{a=1}^A\sum_{g\in\{\mathrm{f},\mathrm{m}\}}\4 \rho_{a,g,k}(t)}
	\end{equation}
	as well as
	\[
		 \hat{\sigma}_k^{\mathrm{MAPappr}}:=\sqrt{
		\frac{1}{T}\sum_{t=1}^T \big(\hat{\lambda}^{\mathrm{MAPappr}}_k(t)-1\big)^2}\,,\quad k\in\{1,\dots,K\}\,,
	\]
	which is simply the sample variance of $\boldsymbol{\hat{\lambda}}^{\mathrm{MAP}}$.
	
	Much faster to derive but also less accurate, we can use a matching of moments approach. Therefore, we have to make the simplifying assumption that deaths\/
	$(N_{a,g,k}(t))_{t\in U}$ are i.i.d.~which can approximatively be obtained by modifying
	the deaths\/ $n_{a,g,k}(t)$	via
	\begin{equation}\label{transform}
		n'_{a,g,k}(t):=\bigg\lfloor\frac{m_{a,g}(T)\4
		q_{a,g}(T)\4 w_{a,g,k}(T)}{m_{a,g}(t)\4
		q_{a,g}(t)\4 w_{a,g,k}(t)}\4
		n_{a,g,k}(t)\bigg\rfloor\,,\quad t\in U\,,
	\end{equation}
	and, correspondingly, $m_{a,g}:=m_{a,g}(T)$, 
	as well as
	$q_{a,g}:=q_{a,g}(T)$
	and 
	$w_{a,g,k}:=w_{a,g,k}(T)$.

Estimates $\hat{q}^{\mathrm{MM}}_{a,g}(t)$
	for death 
	probabilities $q_{a,g}(t)$ can be obtained via 
	minimising mean squared error to death rates which,
	if parameters $\zeta$, $\eta$ and $\kappa$ are previously fixed, can be obtained by  regressing 
	\[
		(F^{\mathrm{Lap}})^{-1}\bigg(\frac{\sum_{k=0}^K n'_{a,g,k}(t)}{m_{a,g}(t)}\bigg)-\kappa_{z_{a,g}}
	\]
	on $\mathcal{T}_{\zeta_{a,g},\eta_{a,g}}(t)$.
	Estimates
	$\hat{u}^{\mathrm{MM}}_{a,g,k},\hat{v}^{\mathrm{MM}}_{a,g,k},\hat{\phi}^{\mathrm{MM}}_{k},\hat{\psi}^{\mathrm{MM}}_{k}$  for parameters $u_{a,g,k},v_{a,g,k},\phi_k,\psi_k$ via
	minimising the mean squared error to death rates which
	again, if parameters $\phi$ and $\psi$ are previously fixed, can be obtained by regressing 
	\[
		\log\frac{n'_{a,g,k}(t)}{m_{a,g}(t)\4 \hat{q}^{\mathrm{MM}}_{a,g}(t)}
	\]
	on $\mathcal{T}_{\phi_k,\psi_k}(t)$.
	Estimates 
	$\hat{w}^{\mathrm{MM}}_{a,g,k}(t)$ 
	are then given by (\ref{eq:WeightFamily}).

Defining unbiased estimators for weights $W^*_{a,g,k}(t):=N_{a,g,k}(t)/(m_{a,g}\4 q_{a,g})$,	as well as
$\overline{W}^{*}_{a,g,k}:=\frac{1}{T}\sum_{t=1}^T{W}^{*}_{a,g,k}(t)$ gives estimator 
	\[
		\widehat{\Sigma}_{a,g,k}^2=\frac{1}{T-1}\sum_{t=1}^T\big({W}^{*}_{a,g,k}(t)-\overline{W}^{*}_{a,g,k}\big)^2\,,
	\]
	 we have
	\[
		\expvbig{\widehat{\Sigma}_{a,g,k}^2}=\varbig{{W}^{*}_{a,g,k}(t)}
		=\frac{w_{a,g,k}}{ m_{a,g}\4 q_{a,g}}+\sigma^2_k\4w_{a,g,k}^2\,.
	\]
	Thus, \emph{matching of moments estimate}
	for $\sigma_k$ can be defined as
	\[
		\hat{\sigma}_{k}^{\mathrm{MM}}:=
		\sqrt{\max\Biggl\lbrace0,\frac{\sum_{a=1}^A\sum_{g\in\{\mathrm{f},\mathrm{m}\}}\Big(\hat{\sigma}_{a,g,k}^2
		-\frac{ w^{\mathrm{MM}}_{a,g,k}(T)}{m_{a,g}\4 q^{\mathrm{MM}}_{a,g}(T)}\Big)}{\sum_{a=1}^A\sum_{g\in\{\mathrm{f},\mathrm{m}\}} (w^{\mathrm{MM}}_{a,g,k}(T))^2}\Biggr\rbrace}
		\,,
	\]
	where $\hat{\sigma}_{a,g,k}^2$ is the estimate corresponding to estimator $\widehat{\Sigma}_{a,g,k}^2$.

	\section{Real world example}\label{sec:real}
	
	\subsection{Prediction of death cause intensities}

As an applied example for estimation in our model, as well as for some further applications,  we take annual death data from Australia for the period 1987 to 2011. We fit our annuity model 
 using the matching of moments approach, as well as the maximum-likelihood approach 
with Markov chain Monte Carlo (MCMC). 
Data source for historical Australian population, categorised by age and gender, is taken from the 
\href{http://www.abs.gov.au/AUSSTATS/abs@.nsf/DetailsPage/3101.0Jun\%202013?OpenDocument}{Australian Bureau
of Statistics} and data for the number of deaths categorised by death cause and divided into eight age categories, i.e., 50--54 years, 
55--59 years, 60--64 years, 65--69 years, 70--74 years, 75--79 years, 80--84 years and 85+ years, denoted by $a_1,\dots,a_8$, respectively, 
for each gender 
is taken from the \href{http://www.aihw.gov.au/deaths/aihw-deaths-data/\#nmd}{AIHW}. The provided death data 
is divided into 19 different death causes---based on the ICD-9 or ICD-10 classification---where we identify the following ten
of them with common non-idiosyncratic risk factors: \textit{\lq certain infectious and parasitic diseases\rq, 
\lq neoplasms\rq, \lq endocrine, nutritional and metabolic diseases\rq, \lq mental and behavioural disorders\rq, 
\lq diseases of the nervous system\rq, \lq circulatory diseases\rq, \lq diseases of the respiratory system\rq, 
\lq diseases of the digestive system\rq,
\lq external causes of injury and poisoning\rq,
\lq diseases of the genitourinary system\rq}. We merge the remaining eight death causes to idiosyncratic risk as 
their individual contributions to overall death counts are small for all categories. 
Data handling needs some care as there was a change in classification of  death data in 1997 as explained at the website of the \href{http://www.abs.gov.au/ausstats/abs@.nsf/Products/3303.0~2007~Appendix~Comparability+of+statistics+over+time+\%28Appendix\%29?OpenDocument}{Australian Bureau of Statistics}.
Australia introduced the tenth revision of the International Classification of Diseases (ICD-10, following ICD-9) in 1997, with a transition period from 1997 to 1998. Within this period, 
comparability factors were produced.
Thus, for the period 1987 to 1996, death counts have to be multiplied by corresponding  
comparability factors.

Trends are considered as described above where cohort and
trend reduction parameters are fixed a priori with values $t_0=1987$, $\kappa_{z_i}=0$,
$\zeta_{a_i,g}=\phi_k=0$ and $\eta_{a_i,g}=\psi_k=\frac{1}{150}$ where the last value 
is based on mean trend reduction in death probabilities with the stated parameter combination. 
For a more advanced modelling of trend reduction see Figure \ref{fig:PDforecast3}.
Thus, within the maximum-likelihood framework, 
we end up with $394$ parameters, with 362 to be optimised, in our annuity model. 
As deterministic numerical optimisation of the likelihood function breaks down due to high dimensionality, we use MCMC in this 
maximum-likelihood setting instead. 
Based on 40\,000 MCMC steps with burn-in period of 10\,000 we are able to derive estimates of all parameters. The execution time of our algorithm is roughly seven hours on a standard computer in \lq R\rq, several parallel MCMC chains 
can be run, each with different starting values.
\begin{figure}[ht]
	\begin{center}
		\includegraphics[width=0.85\textwidth]{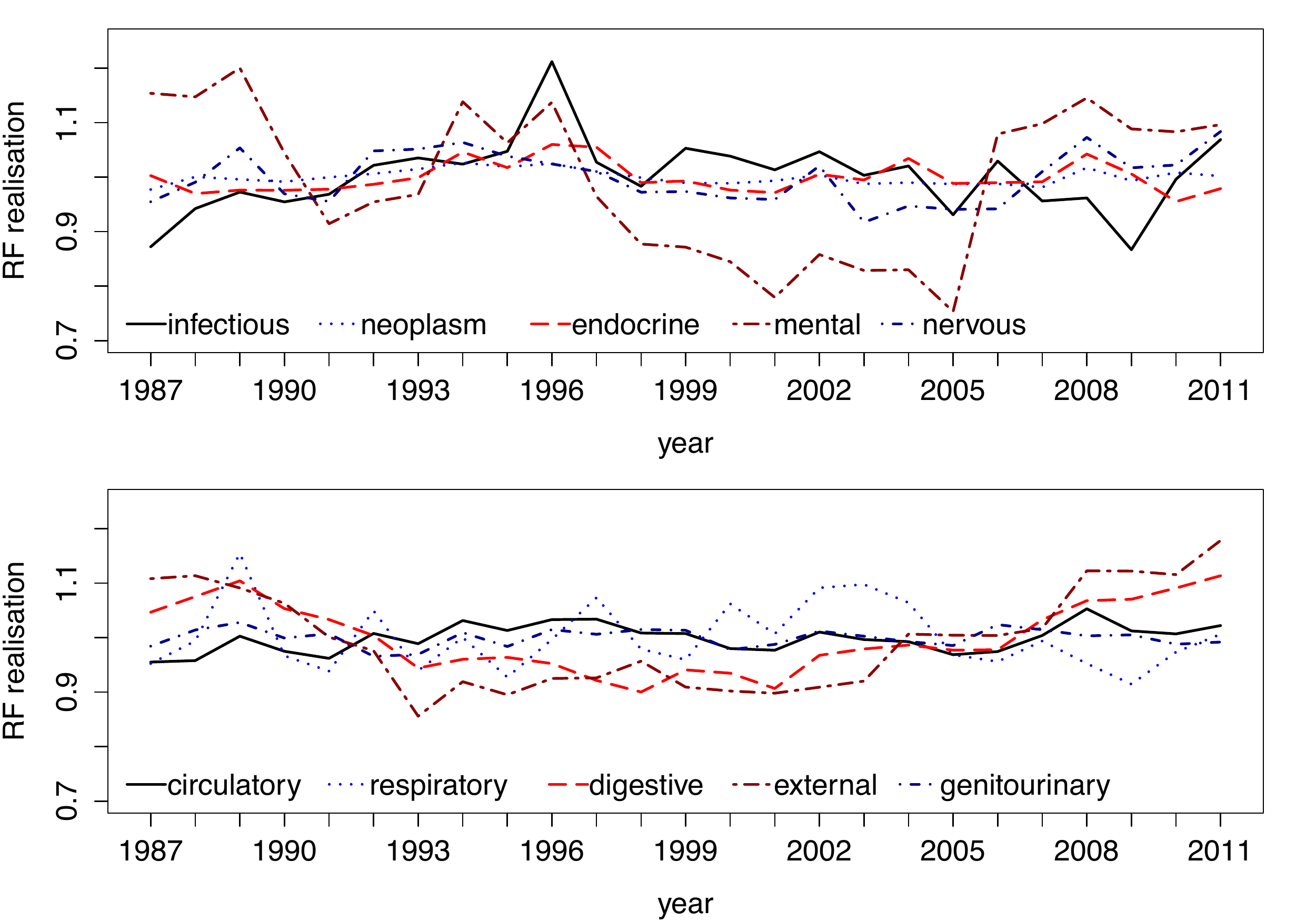}
		\caption[Example Australia: Risk factor realisations 1987--2011.]{Estimated risk factor realisations.}
		\label{fig:RFreal}
	\end{center}
\end{figure}
We can use 
Equation (\ref{lambda_MAP}) to derive approximations for risk factor realisation estimates where all required parameters are taken from the  MCMC estimation. 
Results are shown in Figure \ref{fig:RFreal}. In the top figure we observe a massive jump in the 
risk factor for mental and behavioural disorders between 2005 to 2006 which is mainly driven by 
an unexpectedly high increase in deaths due to dementia which is explained by the ABS as being due to new coding instructions and changes to various Acts, see Chapter 2.4 of the report on \href{http://www.aihw.gov.au/WorkArea/DownloadAsset.aspx?id=10737422943}{Dementia in Australia of the AIHW (2012)}. 

\begin{table}[ht]
	\begin{center}\footnotesize{
		\caption[Example Australia: Estimated and forecasted weights.]{Estimated weights for all death causes with five and\/ 95 percent quantiles in brackets.}
		\setlength{\extrarowheight}{0.5pt}
		\label{tab:deathCauses}
		\begin{tabular}{r|rr|rr|rr|rr}
			\cline{2-9}\rule[-6pt]{0pt}{18pt}
			&\multicolumn{4}{ c | }{male}&\multicolumn{4}{ c  }{female}\\
			\cline{2-9}\rule[-6pt]{0pt}{18pt}
			&\multicolumn{2}{ c | }{60 to 64 years}&\multicolumn{2}{ c | }{80 to 84 years}
			&\multicolumn{2}{ c | }{60 to 64 years}&\multicolumn{2}{ c  }{80 to 84 years}\\
			\cline{2-9}\rule[-6pt]{0pt}{18pt}
			  & 2011 &  2031 \scriptsize{(quant.)} & 2011 &  2031 \scriptsize{(quant.)}
				 & 2011 &  2031 \scriptsize{(quant.)} & 2011 &  2031 \scriptsize{(quant.)}\\
				\hline\rule[10pt]{0pt}{0pt}
		neop. & 0.499& 0.547 $\big(\substack{0.561\\ 0.531}\big)$& 0.324 & 0.378 $\big(\substack{0.392\\ 0.364}\big)$&0.592& 0.648 $\big(\substack{0.662 \\0.629}\big)$&0.263  &0.303 $\big(\substack{0.319\\ 0.288}\big)$\\[0.8ex]
		circ. & 0.228  &0.116 $\big(\substack{0.123 \\0.109}\big)$& 0.325 &0.173 $\big(\substack{0.181 \\0.164}\big)$& 0.140& 0.060 $\big(\substack{0.065 \\0.055}\big)$&  0.342  &0.149 $\big(\substack{0.158\\ 0.140}\big)$\\[0.8ex]
		ext.& 0.056& 0.062 $\big(\substack{0.073 \\0.053}\big)$& 0.026& 0.028 $\big(\substack{0.033\\ 0.024}\big)$& 0.072 &0.069   $\big(\substack{0.078 \\0.060}\big)$&  0.100&  0.126 $\big(\substack{0.139\\ 0.113}\big)$\\[0.8ex]
		resp.& 0.051 & 0.036 $\big(\substack{0.040 \\0.032}\big)$& 0.106  &0.092 $\big(\substack{0.101\\ 0.083}\big)$& 0.038  &0.037 $\big(\substack{0.043\\ 0.032}\big)$& 0.051& 0.068 $\big(\substack{0.074\\ 0.061}\big)$\\[0.8ex]
		endo.&0.044  &0.062 $\big(\substack{ 0.070 \\0.055}\big)$& 0.047&  0.077 $\big(\substack{0.084\\ 0.070}\big)$& 0.036 & 0.051 $\big(\substack{0.060 \\0.043}\big)$&0.054& 0.080 $\big(\substack{0.089\\ 0.071}\big)$\\[0.8ex]
		dig. &0.041 &0.036 $\big(\substack{0.040\\ 0.031}\big)$&  0.027 &0.020 $\big(\substack{0.023\\ 0.018}\big)$& 0.035  &0.032 $\big(\substack{0.038\\ 0.026}\big)$& 0.024 & 0.023 $\big(\substack{0.027 \\0.020}\big)$\\[0.8ex]
		nerv. & 0.029  &0.052 $\big(\substack{0.061\\ 0.045}\big)$& 0.045& 0.061 $\big(\substack{0.068\\ 0.055}\big)$&0.031& 0.024 $\big(\substack{ 0.029\\ 0.020}\big)$& 0.034&0.023 $\big(\substack{0.027\\ 0.020}\big)$\\[0.8ex]
		idio. &0.018 & 0.028 $\big(\substack{0.034\\ 0.023}\big)$&0.015 &0.018 $\big(\substack{0.020\\ 0.016}\big)$& 0.022 & 0.023 $\big(\substack{0.028\\ 0.019}\big)$& 0.023 &0.024 $\big(\substack{0.027 \\0.022}\big)$\\[0.8ex]
		inf. & 0.014 &0.025 $\big(\substack{0.033 \\0.020}\big)$& 0.015  &0.022 $\big(\substack{0.027\\ 0.019}\big)$&  0.014 &0.020  $\big(\substack{0.027 \\0.015}\big)$&  0.017& 0.024 $\big(\substack{0.028\\ 0.020}\big)$\\[0.8ex]
		ment. & 0.013  &0.027 $\big(\substack{0.036 \\0.019}\big)$&  0.041 &0.105 $\big(\substack{0.130\\ 0.078}\big)$& 0.012 &0.032 $\big(\substack{0.046\\ 0.021}\big)$& 0.062 &0.155 $\big(\substack{0.188 \\0.118}\big)$\\[0.8ex]\rule[-7pt]{0pt}{0pt}
		geni. & 0.008& 0.008 $\big(\substack{0.010\\ 0.006}\big)$&  0.028 &0.025 $\big(\substack{0.028 \\0.023}\big)$&  0.009 &0.005 $\big(\substack{0.006 \\0.004}\big)$&0.029 &0.026 $\big(\substack{0.028\\ 0.023}\big)$\\[0.8ex]
\hline
		\end{tabular}}
	\end{center}
\end{table}

Assumption (\ref{eq:WeightFamily}) provides a joint forecast of all death cause intensities, i.e.~weights, simultaneously---in contrast to standard procedures where projections are made for each death cause separately.
As already presumed in Figure \ref{fig:death_causes} in the introduction, our model 
observes major shifts in weights of certain death causes over previous years as shown in 
Tables \ref{tab:deathCauses} and \ref{tab:LeadingCauses}. This table lists weights $w_{\textrm{a},\textrm{g},k}(t)$ for all death causes estimated for year 2011, as well as forecasted for 
2031 using (\ref{eq:WeightFamily}) with MCMC mean estimates for ages $60$ to $64$ years (left) and $80$ to $84$ years (right). Our model forecasts suggest that if these trends in weight changes persist, then the future gives a whole new picture of mortality. First, deaths due to circulatory diseases are expected to decrease whilst neoplasms will become 
the leading death cause over most age categories.
Moreover, deaths due to mental and behavioural disorders are expected to rise massively for older ages.
 This observation nicely illustrates the serial dependence, amongst different death causes captured by our model. This potential increase in deaths 
due to mental and behavioural disorders for older ages will have a massive 
impact on social systems as, typically, such patients need long-term geriatric care. High uncertainty in forecasted weights
is reflected by wide confidence intervals (values in brackets) for the risk factor of mental and behavioural disorders. These confidence intervals are derived from corresponding MCMC chains 
and, therefore, solely reflect uncertainty associated with parameter estimation. Note that results for estimated trends depend on the length of the data period as short-term trends might not coincide with mid- to long-term trends.

\begin{table}[ht]
	\begin{center}\footnotesize{
		\caption[Example Australia: Estimated and forecasted leading weights for males and females.]{Leading death causes with weights in brackets.}
		\setlength{\extrarowheight}{0.5pt}
		\label{tab:LeadingCauses}
		\begin{tabular}{rr|rr|rr}
						\cline{3-6}\rule[-6pt]{0pt}{18pt}
				&&\multicolumn{2}{ c }{male}&\multicolumn{2}{ c }{female}\\
			\cline{3-6}\rule[-6pt]{0pt}{18pt}
			  && 2011 &  2051 & 2011 &  2051 \\
			\hline\rule[10pt]{0pt}{0pt}
		&1.& neoplasms~(0.469) &neoplasms~(0.474)& neoplasms~(0.603) &neoplasms~(0.581)\\
		55--59 years& 2.& circulatory (0.222)&  infectious (0.092)& circulatory (0.112)& nervous (0.077)\\\rule[-5pt]{0pt}{0pt}
		& 3.& external (0.085) & external (0.083)& respiratory (0.058) & not elsewhere (0.068)\\
\hline\rule[10pt]{0pt}{0pt}
		&1.& neoplasms~(0.505) &neoplasms~(0.575)& neoplasms~(0.551) &neoplasms~(0.609)\\
		65--69 years& 2.& circulatory (0.226)&  endocrine (0.082)& circulatory (0.162)&  mental (0.112)\\\rule[-5pt]{0pt}{0pt}
		& 3.& respiratory (0.072) & mental (0.075)& respiratory (0.083) & nervous (0.065)\\
\hline\rule[10pt]{0pt}{0pt}
		&1.& neoplasms~(0.405) &neoplasms~(0.466)& neoplasms~(0.365) &neoplasms~(0.378)\\
		75--79 years& 2.& circulatory (0.277)& mental (0.185)& circulatory (0.271)&  mental (0.245)\\\rule[-5pt]{0pt}{0pt}
		& 3.& respiratory (0.100) &  endocrine (0.098)& respiratory (0.103) & respiratory (0.108)\\
\hline\rule[10pt]{0pt}{0pt}
		&1.& circulatory~(0.395) &mental~(0.329)& circulatory~(0.441) &mental~(0.503)\\
		85+ years& 2.& neoplasms (0.217)&  neoplasms (0.216)& neoplasms (0.131)& circulatory (0.092)\\\rule[-5pt]{0pt}{0pt}
		& 3.& respiratory (0.115) & circulatory (0.133)& mental (0.101) & neoplasms (0.090)\\
\hline
		\end{tabular}}
	\end{center}
\end{table}

\subsection{Scenario analysis}
\begin{figure}[ht]
	\begin{center}
		\includegraphics[width=0.75\textwidth]{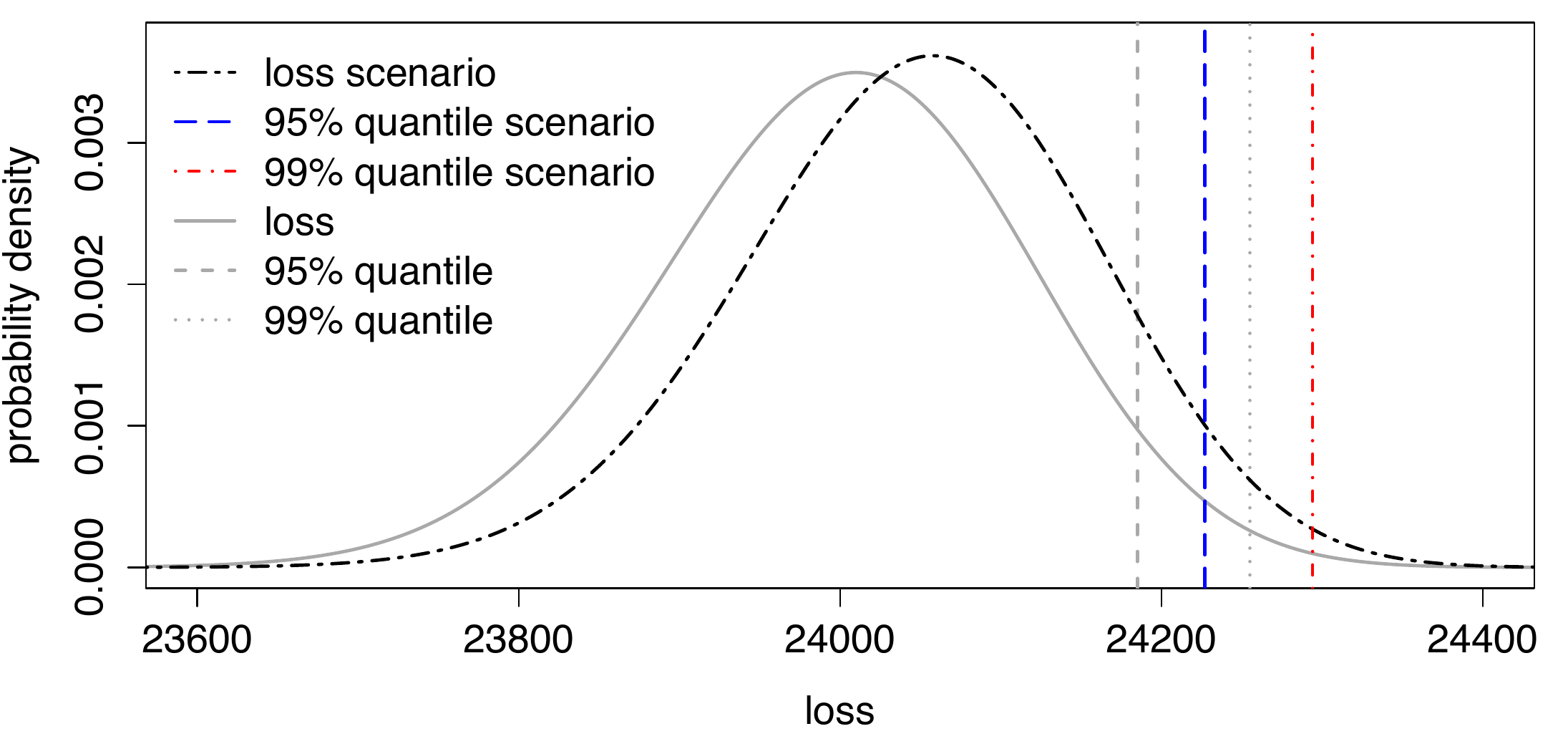}
		\caption[Example Australia: Loss distribution for simple portfolio with scenario.]{Loss distributions of\/ $L$ and\/ $L^{\mathrm{scen}}$ with\/ 95 and\/ 99 percent quantiles.}
		\label{fig:scen}
	\end{center}
\end{figure}

As an explanatory example of our annuity model, assume $m=1\,600$ policyholders which distribute uniformly over all age categories and genders, i.e., each age
category contains 100 policyholders with corresponding death probabilities, as well as weights as previously estimated and 
forecasted for 2012.
Annuities $X_i=Y_i$ for all $i\in\{1,\dots,m\}$ are paid annually and take deterministic values in $\{11,\dots,20\}$ such that ten policyholders in 
each age and gender category share equally high payments. 
We now want to analyse the scenario, 
indexed by \lq scen\rq , that deaths due to neoplasms are reduced by 25 percent in 2012 over all age categories. In that case, we can estimate the realisation of risk factor 
for neoplasms, see (\ref{lambda_MAP}), which takes an estimated value of $0.7991$. 
Running our annuity model with this risk factor realisation being fixed, we end up 
with a loss distribution $L^\mathrm{scen}$ where deaths due to neoplasms have decreased.
Figure \ref{fig:scen} then shows probability distributions of traditional loss $L$ 
without scenario, as well as of scenario loss  $L^{\mathrm{neo}}$ with corresponding 95 percent and 99 percent 
quantiles.
We observe that a reduction of 25 percent in cancer death rates leads to a 
remarkable shift in quantiles of the loss distribution as fewer people die and, thus, 
more annuity payments have to be made.

\subsection{Forecasting death rates and comparison with the Lee--Carter model}
We  can compare out-of-sample
forecasts of death rates from our model to forecasts obtained by other mortality models. In this paper, we choose the traditional Lee--Carter model as a proxy, see, for example, Kainhofer, Predota and Schmock
\cite[Section 4.5.1]{AVOE_annuity_table}, as our model is conceptionally based on a similar idea and as it is widely used in practice.  
Henceforth, we set $Y_j(t):=1$ for all people $j\in M_{a,g}(t)$ with $M_{a,g}(t)=\{1,\dots,m_{a,g}(T)\}$ and $t\in\{T+1,\dots,S\}$. The simplifying assumption of a constant population is
conservative as population tends to increase and, therefore, statistical fluctuations are over-estimated.
To show a further application of our model we compare out-of-sample forecasts from our model to forecasts obtained by the traditional Lee--Carter model. Given the number of living people $m_{a,g}(t)$, as well as annual deaths $n_{a,g}(t):=\sum_{k=0}^K n_{a,g,k}(t)$, the Lee--Carter approach models logarithmic death rates
$\log { r}_{a,g}(t):=\log n_{a,g}(t)-\log m_{a,g}(t)$
in the form
$\log {r}_{a,g}(t)=\mu_{a,g}+\tau_t\4 \nu_{a,g}+\varepsilon_{a,g,t}$
with independent normal error terms $\varepsilon_{a,g,t}$ with mean zero and 
common time-specific components $\tau_t$. Using suitable 
normalisations, estimates for these components can be derived via method of moments and singular value decompositions,  see, for example, Kainhofer, Predota and Schmock
\cite[Section 4.5.1]{AVOE_annuity_table}. Forecasts may then be obtained by using 
auto-regressive models for $\tau$. 
Conversely,
using our model it is straight-forward to \emph{forecast death rates}
and to give corresponding confidence intervals via setting $Y_j(t):=1$ for all people $j\in M_{a,g}(t)$ with $M_{a,g}(t)=\{1,\dots,m_{a,g}(T)\}$ and $t\in\{T+1,\dots,S\}$. The assumption of constant population for forecasts in Australia is 
conservative as population tends to increase and, therefore, statistical fluctuations are over-estimated. As an alternative, 
more sophisticated models for population forecasts can be used.

Then, for an estimate $\hat{\boldsymbol{\theta}}$ of parameter vector 
$\boldsymbol{\theta}$ run 
our annuity model with parameters 
forecasted, see (\ref{eq:PDFamily}) and (\ref{eq:WeightFamily}). 
We then obtain the distribution of the total number 
of deaths $S_{a,g}(t)$ given $\hat{\boldsymbol{\theta}}$ and, thus,
forecasted death rate $\hat{r}_{a,g}(t)$ is given by
\[
	\Prob\Big(\hat{r}_{a,g}(t)=\frac{n}{m_{a,g}(T)}\Big)
	=\Prob(S_{a,g}(t)=n)\,, \quad n\in\na_0\,.
\]
Uncertainty in the form of confidence intervals represent 
statistical fluctuations, as well as random changes in risk factors. Additionally, using results obtained 
by Markov chain Monte Carlo (MCMC) it is even possible to 
incorporate parameter uncertainty into predictions.
To account for an increase in uncertainty for forecasts we suggest to assume increasing risk factor variances for forecasts, e.g., $\tilde\sigma^2_k(t)=\sigma^2_k\4 (1+d\4(t-T))^2$ with $d\geq 0$. 
A motivation for this approach with $k=1$ is the following: A major source of uncertainty for forecasts lies in an 
unexpected deviation from the estimated trend for death probabilities. We may therefore assume that rather than being deterministic, forecasted 
values $q_{a,g}(t)$ are beta distributed (now denoted by $Q_{a,g}(t)$) with $\expv{Q_{a,g}(t)}=q_{a,g}(t)$ and 
variance $\sigma^2_{Q,a,g}(t)$ which is increasing in time. Then, given independence amongst risk factor $\Lambda_1$ and 
$Q_{a,g}(t)$, we may assume that there exists a future point $t_0$ such that 
\[
	\sigma^2_{Q,a,g}(t_0)=\frac{q_{a,g}(t_0)\4 (1- q_{a,g}(t_0))}{1/\sigma^2_1+1}\,.
\]
In that case, $Q_{a,g}(t)\4 \Lambda_1$ is again gamma distributed with mean one and increased variance 
$q_{a,g}(t_0)\4\sigma^2_1$ (instead of $q^2_{a,g}(t_0)\4\sigma^2_1$ for the deterministic case). 
Henceforth, it seems reasonable to stay within the family of gamma distributions for forecasts and just adapt 
variances over time. Of course, families for these variances for gamma distributions can be changed arbitrarily
and may be selected via classical information criteria.

\begin{figure}[ht]
	\begin{center}
		\captionsetup{aboveskip=0.2\normalbaselineskip}
		\includegraphics[width=0.75\textwidth]{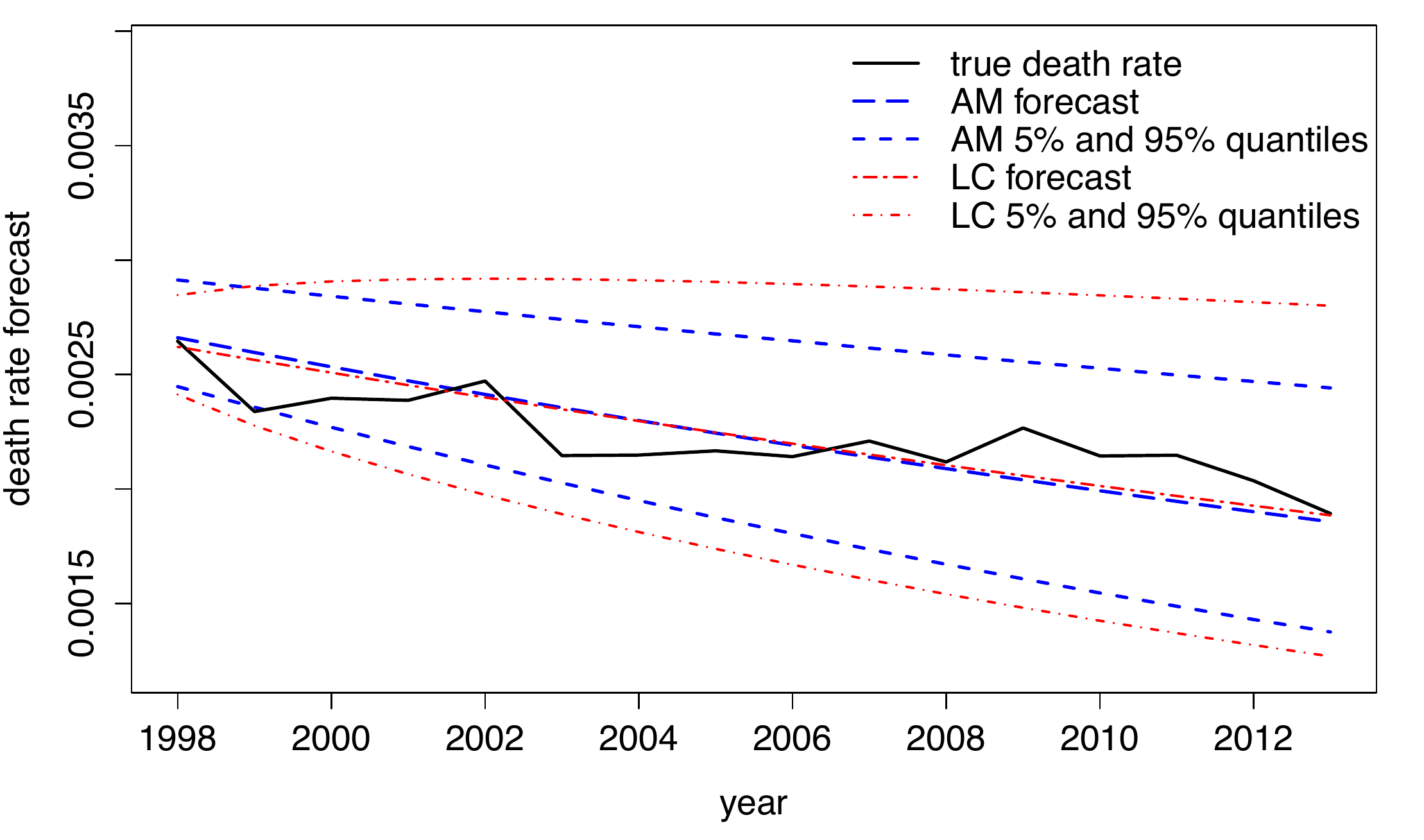}
		\caption[Example Australia: True versus forecasted death rates 2002--2011.]{Forecasted death rates using our annuity model (AM) and the Lee--Carter model (LC).}
		\label{fig:PDforecast}
	\end{center}
\end{figure}

Using	in-sample data, $d$ can be estimated using (\ref{MLE_likelihood}) with all other parameters being fixed. Using Australian death and population data for the years 1963 to 1997 we estimate model parameters via MCMC in our annuity model with one common stochastic risk factor having constant weight one. In average, i.e., for various forecasting periods and starting at different dates, parameter $d$ takes the value $0.22$ in our example.
Using fixed trend parameters as above, and using the mean of 30\,000 MCMC samples, we forecast death rates and corresponding confidence intervals out of sample
for the period 1998 to 2013. We can then compare these results to realised death rates within the stated period and 
to forecasts obtained by the Lee--Carter model which is shown in Figure \ref{fig:PDforecast} for females aged $50$ to $54$ years. 
We observe that true death rates mostly fall in the 90 percent confidence band for both procedures.
Moreover, Lee--Carter forecasts lead to wider spreads of quantiles in the future whilst our model suggests a more moderate increase in uncertainty. Taking various parameter samples from the MCMC chain and deriving quantiles for death rates, we can extract contributions of parameter 
uncertainty in our model coming from posterior distributions of parameters.

\subsection{Forecasting death probabilities}\label{sec:forecaDP}

\begin{figure}[ht]
	\begin{center}
		\captionsetup{aboveskip=0.2\normalbaselineskip}
		\includegraphics[width=1\textwidth]{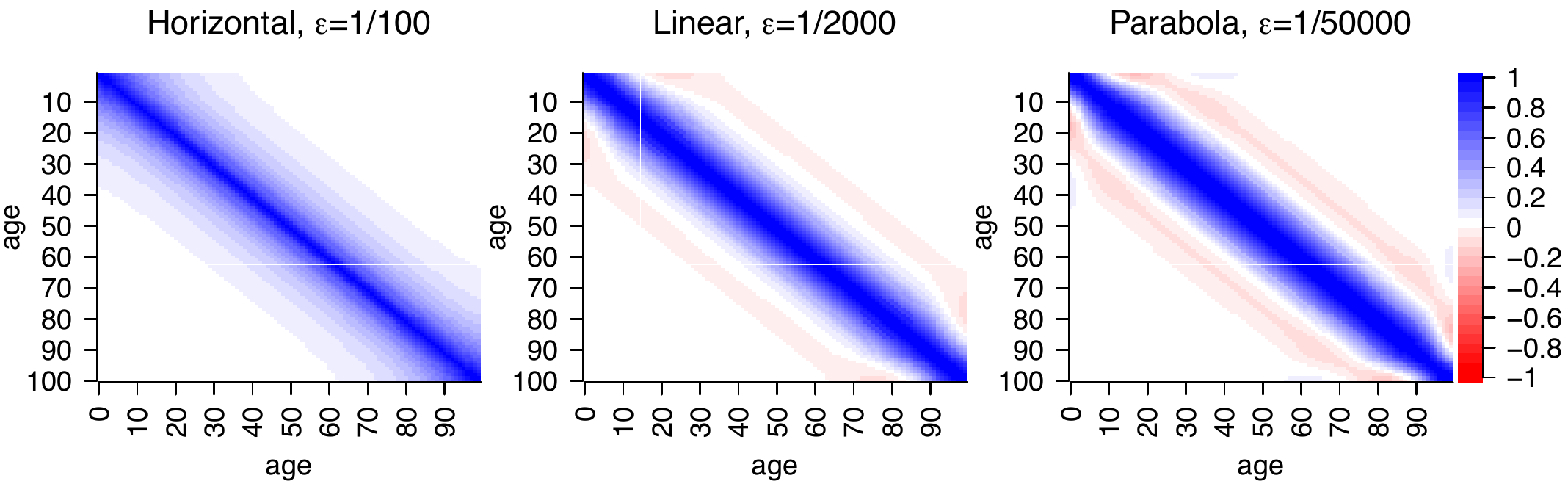}
		\caption[Correlation structure for Gaussian priors.]{Correlation structure of Gaussian priors with penalisation for deviation from ordinate (left), straight line (middle) and parabola (right).}
		\label{fig:corr}
	\end{center}
\end{figure}

Forecasting death probabilities within our annuity model is straight forward using 
(\ref{eq:PDFamily}). In the special case with just idiosyncratic risk, i.e., $K=0$, death indicators can be assumed to be Bernoulli distributed instead of being Poisson distributed in which case we may write the likelihood function in the form
\[
		\ell^{\mathrm{B}}(\boldsymbol{n}\conditioned\alpha,\beta,\zeta,\eta,\kappa)=
		\prod_{t=1}^T \prod_{a=1}^A\prod_{g\in\{\mathrm{f},\mathrm{m}\}} \binom {m_{a,g}(t)} {n_{a,g,0}(t)}
		q_{a,g}(t)^{n_{a,g,0}(t)}\4 (1-q_{a,g}(t))^{m_{a,g}(t)-n_{a,g,0}(t)}\,,
\]
with $0\leq n_{a,g,0}(t)\leq m_{a,g}(t)$.  
Due to possible overfitting, derived estimates may not be sufficiently smooth across age categories $a\in\{1,\dots,A\}$. Therefore, if we switch to a Bayesian setting, we may use regularisation via prior distributions. To guarantee smooth results and a sufficient stochastic foundation, we suggest the usage of Gaussian priors with mean zero and a specific correlation structure, 
i.e., $\pi(\alpha,\beta,\zeta,\eta,\kappa)=\pi(\alpha)\4\pi(\beta)\4\pi(\zeta)\4\pi(\eta)\4\pi(\kappa)$ with
	\begin{equation}\label{prior_TV}
		\log\pi(\alpha):=	-c_\alpha\4 \sum_{g\in\{\mathrm{f},\mathrm{m}\}}\bigg(
			\sum_{a=1}^{A-1}  (\alpha_{a,g}-\alpha_{a+1,g})^2
			+\varepsilon_\alpha\4 \sum_{a=1}^{A} \alpha_{a,g}^2\bigg)+\log(d_\alpha)\,,\quad c_\alpha,d_\alpha,\varepsilon_\alpha>0\,,
	\end{equation}
	and correspondingly for $\beta$, $\zeta$, $\eta$ and $\kappa$. Parameters $c_\alpha$ (correspondingly 
	for $\beta$, $\zeta$, $\eta$ and $\kappa$) is a scaling parameters and directly associated with the variance of Gaussian priors while 
	normalisation-parameter $d_\alpha$ guarantees that $\pi(\alpha)$ is a proper Gaussian density. 
	Penalty-parameter $\varepsilon_\alpha$ scales the correlation amongst neighbour parameters in the sense that the lower it gets, the higher the correlation. The more we increase $c_\alpha$ 
	the stronger the influence of,  or the believe in  the prior distribution. This particular prior distribution penalises 
 deviations from the ordinate which is 
	a mild conceptual shortcoming as this does not accurately reflect our prior believes. Nevertheless, it yields good results and 
	allows a direct analysis of prior variances and covariances of parameters. Setting  $\varepsilon_\alpha=0$ gives an improper 
	prior with uniformly distributed (on $\re$) marginals such that we gain that there is no prior believe in expectations of parameters but, simultaneously, lose the presence of  variance-covariance-matrices and asymptotically get perfect positive correlation across parameters of different ages. However, setting $\varepsilon_\alpha=0$ very often gives better fits of estimates.
	 An optimal choice of regularisation parameters $c_\alpha,c_\beta,c_\zeta,c_\eta$ and $c_{\kappa}$ can be obtained by cross-validation. Furthermore, we set $\varepsilon_\alpha=\varepsilon_\beta=10^{-2}$ and 
	$\varepsilon_\zeta=\varepsilon_\eta=\varepsilon_\kappa=10^{-4}$
	as this yields a  suitable prior correlation structure which decreases with higher age differences and which is always positive, see the left plot in Figure \ref{fig:corr}.
	 
	 There exist many other reasonable choices for Gaussian prior distributions. For example, replacing graduation terms 
	  $(\alpha_{a,g}-\alpha_{a+1,g})^2$ in (\ref{prior_TV}) by higher order differences  
		of the form $\big(\sum_{\nu=0}^k(-1)^\nu {k\choose\nu} \alpha_{a,g+\nu}\big)^2$
	 yields a penalisation for deviations from a straight line with $k=2$, see middle plot in Figure \ref{fig:corr}, or from a parabola with $k=3$, see right plot in Figure \ref{fig:corr}.
	  The usage of higher order differences for graduation of statistical estimates goes back to
		the Whittaker--Henderson method. Taking $k=2,3$ unfortunately yields negative prior correlations 
	  amongst certain parameters which is why we do not recommend their use. Of course, there exist many further possible 
	  choices for prior distributions.
	 
\begin{figure}[ht]
	\begin{center}
		\captionsetup{aboveskip=0.2\normalbaselineskip}
		\includegraphics[width=1\textwidth]{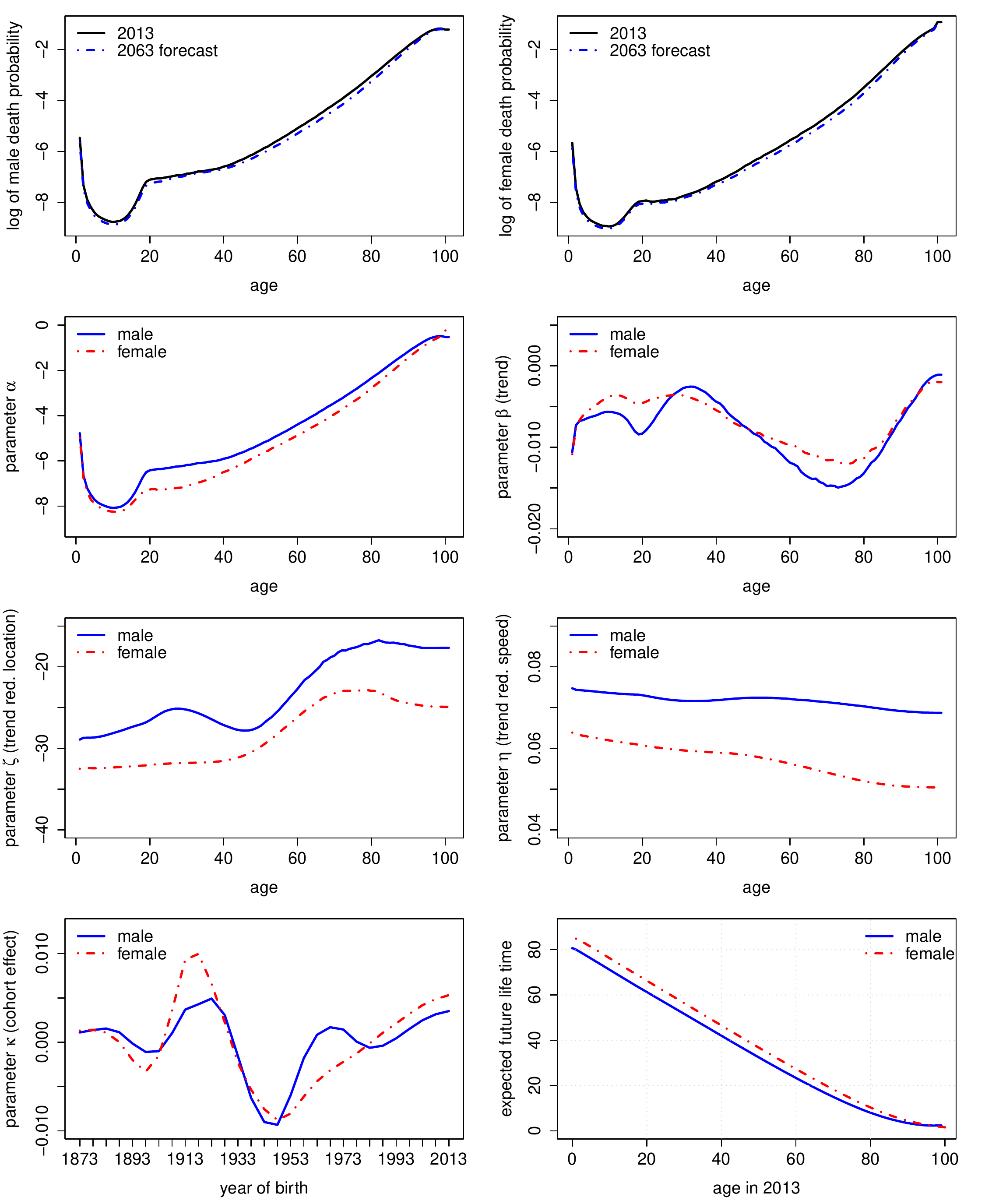}
		\caption{Logarithm of death probabilities (top) for 2013 in Australia and  forecasts for\/ 2063 based on  data from\/ 1971 to\/ 2013, as well as corresponding parameters $\alpha,\beta,\zeta,\eta$ and
		 $\kappa$, as well as expected future lifetime in 2013.}
		\label{fig:PDforecast3}
	\end{center}
\end{figure} 

Results for Australian data 
from 1971 to 2013 with $t_0=2013$ are given in Figure \ref{fig:PDforecast3}. Using MCMC we 
derive estimates for logarithmic death probabilities $\log q_{a,g}(t)$
with corresponding forecasts, mortality trends  $\beta_{a,g}$, as well as trend reduction parameters $\zeta_{a,g},\eta_{a,g}$ and cohort effects $\kappa_{z_{a,g}}$. Recall that $\zeta_{a,g}$ yields the point in time when 
trend acceleration transits into trend reduction. Values close zero indicate that trend reduction 
is about to start whilst negative values indicate that trend reduction is already in place. 
We observe negligible parameter uncertainty due to a long period of data. Further, 
regularisation parameters obtained by cross-validation are given by  $c_\alpha=1\,300$, $c_\beta=c_\eta=30\,000\4 c_\alpha$, $c_\zeta=c_\alpha/20$ and $c_\kappa=1\,000\4c_\alpha$.

We can draw some immediate conclusions. Firstly, we see 
an overall improvement in mortality over all ages where the trend is particularly strong
for young ages and ages between 50  and 80 whereas the trend vanishes towards the age of 100, maybe implying a natural barrier for life expectancy. Due to sparse data the latter conclusion should be treated with the utmost caution.
 Furthermore, we see the classical 
hump of increased mortality driven by accidents around the age of 20 which is more 
developed for males.

Secondly, estimates for $\zeta_{a,g}$ suggest that trend acceleration switched to trend reduction 
throughout the past 20 to 30 years, with a similar shape for males and females. Estimates for 
$\eta_{a,g}$ show that the speed of trend reduction is unexpectedly high, even stronger for males. It should be noted that estimates 
for $\zeta_{a,g}$ and $\eta_{a,g}$ are sensitive to penalty-parameters $\varepsilon_\zeta,\varepsilon_\eta$. Estimates for $\kappa_{z_{a,g}}$ show that the cohort effect is particularly strong (in the sense of increased mortality) for the generation born around 1915---probably 
associated with World War II---and particularly weak for the generation born around 1945.
 

Henceforth, based on forecasts for death probabilities, expected future life time can be estimated. To be consistent concerning longevity risk, mortality trends have to 
be included as a 60-year-old today will probably not 
have as good medication as a 60-year-old in several decades. However, it seems that this is not 
the standard approach in the literature.
Based on the definitions above, expected (curtate) future life time of a person at date $T$ is given by
\begin{equation}\label{lifeExp}
	e_{a,g}(T)=\expv{K_{a,g}(T)}=\sum_{k=1}^\infty 
	{}_k p_{a,g}(T)
\end{equation}
where survival probabilities over $k\in\na$ years are given by
${}_k p_{a,g}(T):=\prod_{j=0}^{k-1} \big(1-q_{a+j,g}(T+j)\big)$
and where $K_{a,g}(T)$ denotes the number of completed future years lived by 
a person of particular age and gender at time $T$. In Australia we get a life expectancy of roughly $80.7$ years for males and $84.9$ for females born in 2013. Thus, 
comparing these numbers to a press release from October 2014 from the \href{http://www.abs.gov.au/ausstats\%5Cabs@.nsf/mediareleasesbyCatalogue/F95E5F868D7CCA48CA25750B0016B8D8?Opendocument
}{Australian Bureau of Statistics} saying that 
\lq Aussie men now expected to live past 80\rq, we get similar results 
whereas our forecasts are slightly higher due to the consideration of
mortality trends. It has to be noted that our results just show a modest increase in life expectancy (compared to the ABS press release) which is mainly 
due to a clear trend reduction, i.e., high values $\eta$, in mortality throughout the past ten years.
If we do not believe in such a strong trend reduction, we may use a longer data history or a priori fix trend reduction parameters to a lower level.
For example if we set $\zeta_{a,g}=1972$, i.e., we assume trend reduction from the beginning of our data set, estimation gets more stable and 
parameter $\eta_{a,g}$ takes levels around $0.005$ for males and $0.01$ for females. In that case, due to negligible trend reduction, our model 
states that Australian men, born in 2013, are 
expected to live roughly $91$ years and, thus, almost closing the gap to women who are also expected to live $91$ years. 

As a second example, we consider Austrian data
from 1965 to 2014 with $t_0=2014$ and apply MCMC, again. In contrast to the Australian example, we assume $\kappa=0$, i.e.~no cohort effects, and set $\varepsilon_\alpha=\varepsilon_\beta=\varepsilon_\zeta=\varepsilon_\eta=0$. Setting 
 $\kappa=0$ drastically reduces execution times, gives more stable estimates across time and, thus, this assumption is particularly useful  for deriving long-term forecasts and life expectancy. 
 Assuming $\varepsilon_\alpha=\varepsilon_\beta=\varepsilon_\zeta=\varepsilon_\eta=0$ shows better fits of death probabilities across different ages. 
Regularisation parameters obtained by cross-validation are given by  $c_\alpha=150$, $c_\beta=2\4 \cdot 10^6$, $c_\zeta=10$ and $c_\eta=2\4 \cdot 10^6$.
The results, see {fig:PDforecast4}, show
an overall improvement in mortality over all ages where the trend is particularly strong
for young ages whereas the trend vanishes towards the age of 100, maybe implying a natural barrier for life expectancy. Due to sparse data the latter conclusion should be treated with the utmost caution.
 Furthermore, we see the classical
hump of increased mortality around the age of 20.
Estimates for $\zeta_{a,g}$ suggest that trend acceleration switched to trend reduction
throughout the past 20 to 30 years. Estimates for
$\eta_{a,g}$ show that the speed of trend reduction is high, even stronger for males over most ages. 
\begin{figure}[ht]
	\begin{center}
		\captionsetup{aboveskip=0.2\normalbaselineskip}
		\includegraphics[width=1\textwidth]{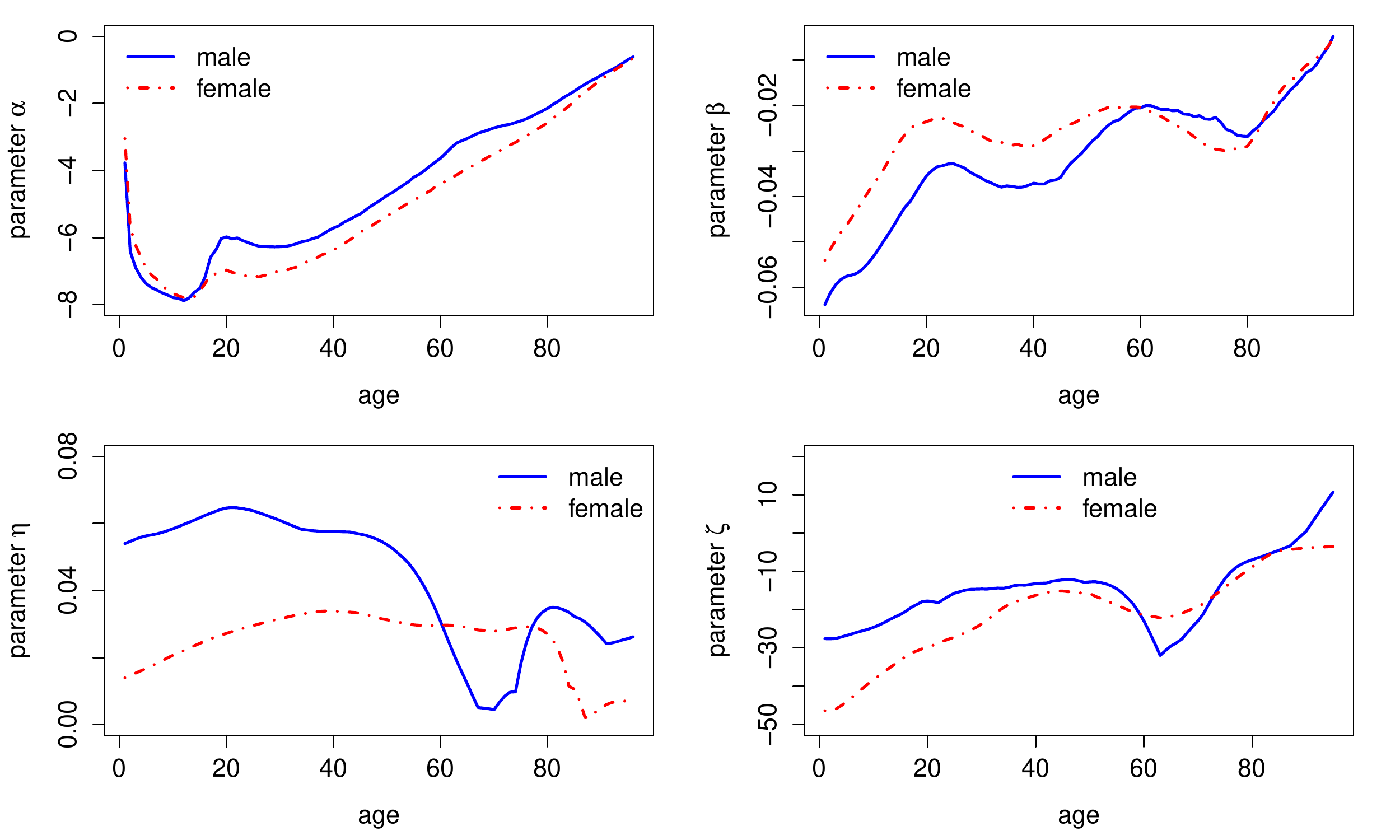}
		\caption{Parameter estimates for $\alpha,\beta,\zeta$ and $\eta$.}
		\label{fig:PDforecast4}
	\end{center}
\end{figure} 

Henceforth, based on forecasts for death probabilities, expected future life time can be estimated. To be consistent concerning longevity risk, mortality trends have to
be included as a 60-year-old today will probably not
have as good medication as a 60-year-old in several decades.
In Austria we get a life expectancy of roughly $84.9$ years for males and $86.9$ for females born in 2014, i.e.~significantly above official figures. For 60-year-old males and females we get  expected future life times of $23.8$ years and $26.1$ years, respectively.

Using estimated values of death probabilities, weights and risk factor variances, as well as population 
forecasts (which are usually freely available at official statistical bureaus), 
our annuity model can be used to derive confidence bands for absolute numbers 
of deaths due to certain causes setting $Y_i=1$. Using different MCMC parameter samples and using
the parameter $\delta$ as described above, parameter risk and uncertainty in forecasts can be incorporated, respectively.

\subsection*{Mortality risk, longevity risk and Solvency II application}
In light of the previous section, life tables can be projected into the future and, thus, it is straightforward to derive best estimate  liabilities (BEL) of annuities and life insurance contracts. The possibility that death probabilities differ from an expected curve, i.e.~estimated parameters do no longer reflect the best estimate and have to  be changed, contributes to mortality or longevity risk, when risk is measured over a one year time horizon as in Solvency II and the duration of in-force insurance contracts exceeds this time horizon. In our model, this risk can be captured by considering various MCMC samples $(\hat{\theta}^h)_{h= 1,\dots, m}$, yielding distributions of BELs. For example, taking $D(T,T+t)$ as the discount curve from time $T+t$ back to $T$ and choosing an MCMC sample $\hat{\theta}^h$ of parameters to calculate death probabilities 
$q^h_{a+t,g}(T+t)$ and survival probabilities ${p}^h_{a,g}(T)$ at age $a$ with gender $g$, the BEL for a term life insurance contract which pays $1$ unit at the end of the year of death within the contract term of $d$ years is given by
\begin{equation}\label{BEL_life_insurance}
	A_{a,g}^T\big(\hat{\theta}^h\big)=D(T,T+1)\4 q^h_{a,g}(T)+\sum_{t=1}^d D(T,T+t+1)\cdot {}_{t} {p}^h_{a,g}(T) \4 q^h_{a+t,g}(T+t)\,.
\end{equation}
In a next step, this approach can be used as a building block for (partial) internal models to calculate basic solvency capital requirements (BSCR) for biometric underwriting risk under Solvency II, as illustrated in the following example.

Consider an insurance portfolio at time $0$ with $n\in\na$ whole life insurance policies with lump sum payments $C_i>0$, for $i=1,\dots,n$, upon death at the end of the year. Assume that all assets are invested in an EU government bond (risk free under the standard model of the Solvency II directive) with maturity $1$, nominal $A_0$  and coupon rate $c>-1$. Furthermore, assume that we are only considering mortality risk and ignore profit sharing, lapse, costs, reinsurance, deferred taxes, other assets and other liabilities, as well as the risk margin. Note that in this case, basic own funds, denoted by $\mathrm{BOF}_t$, are given by market value of assets minus BEL at time $t$, respectively. Then, the BSCR at time $0$ is given by the 99.5\% quantile of the change in basic own funds over the period $[0,1]$, denoted by $\Delta\mathrm{BOF}_1$, which can be derived by, see (\ref{BEL_life_insurance}),
\[
	\begin{split}
	\Delta\mathrm{BOF}_1&=\mathrm{BOF}_0- D(0,1)\4\mathrm{BOF}_1=
		A_0\big(1- D(0,1)\4(1+c)\big)-\sum_{i=1}^n C_i\4A_{a,g}^{0}\big(\hat{\theta}\big)\\
		&\quad+\frac{D(0,1)}{m}\sum_{h=1}^m \bigg(\sum_{i=1}^n C_i\4 A_{a+1,g}^{1}\big(\hat{\theta}^h\big)+\sum_{i=1}^n\sum_{j=1}^{N_i^h}C_i\4 \big(1-A_{a+1,g}^{1}\big(\hat{\theta}^h\big)\big)\bigg)\,.
	\end{split}
\]
where $\hat \theta:=\frac{1}{m}\sum_{h=1}^m \hat{\theta}^h$ and where $N^h_1,\dots,N^h_n$ are independent and Poisson distributed with $\expv{N_i^h}=q_{a_i,g_i}^h(0)$ with policyholder $i$ belonging to age group $a_i$ and of gender $g_i$. The distribution of the last sum above can be derived efficiently by Panjer recursion. This example does not require a consideration of market risk and it nicely illustrates how mortality risk splits into a part associated with statistical fluctuation (experience variance: Panjer recursion) and into a part with long-term impact (change in assumptions: MCMC). Note that by mixing $N_i$ with common stochastic risk factors, we may include other biometric risks such as morbidity. 

\begin{figure}[ht]
	\begin{center}
		\captionsetup{aboveskip=0.2\normalbaselineskip}
		\includegraphics[width=0.6\textwidth]{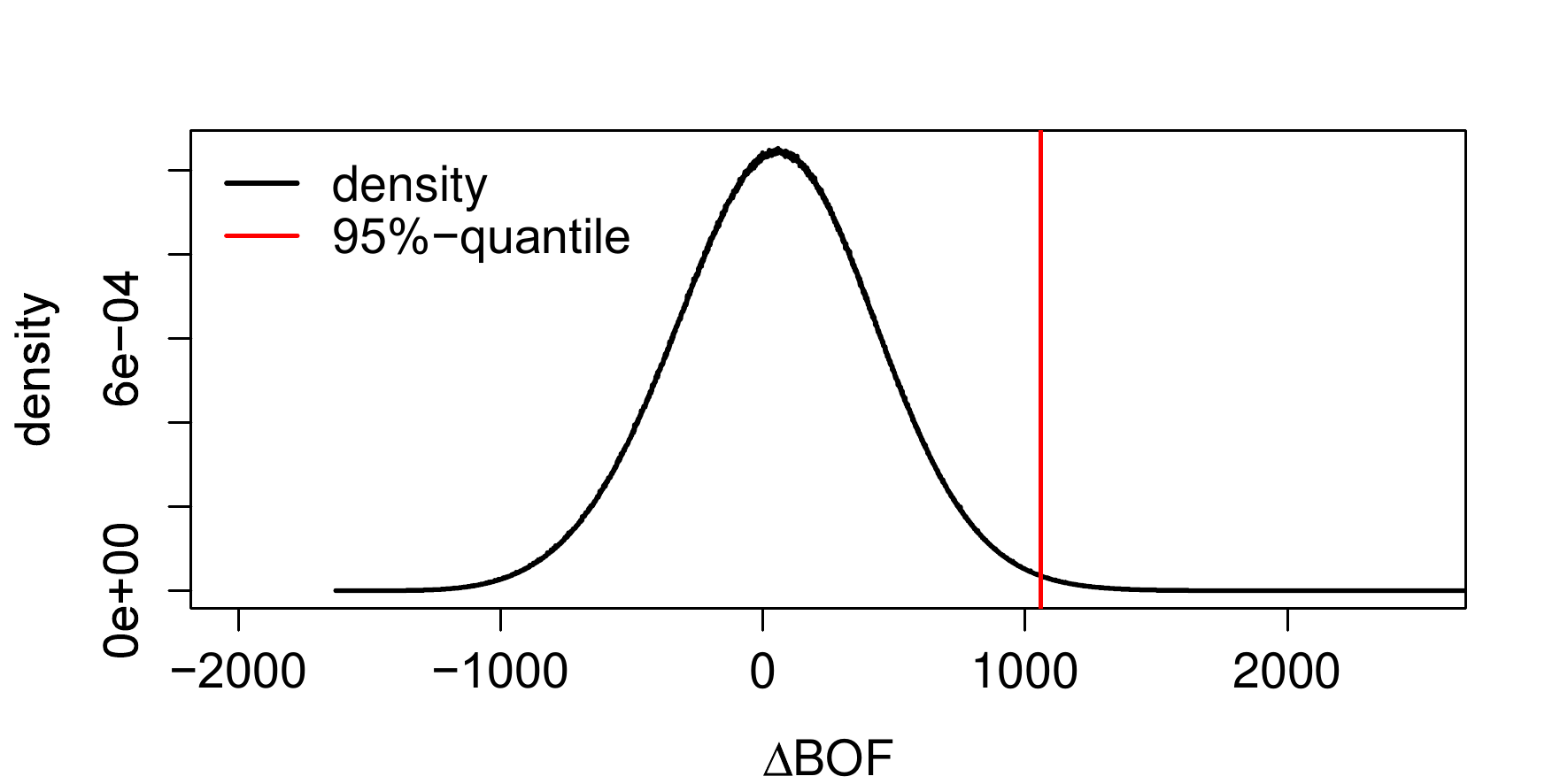}
		\caption{Change in basic own funds, in thousand.}
		\label{fig:changeBOF}
	\end{center}
\end{figure}

Consider a portfolio with 100  males and females at each age between 20 and 60 years, each having a 40-year term life insurance, issued in 2014, which provides a lump sum payment between 10\,000 und 200\,000 (randomly chosen for each policyholder) if death occurs within these 40 years. Using MCMC samples and estimates based on the Austrian data
from 1965 to 2014 as given in the previous section, we may derive the change in basic own funds from 2014 to 2015 by (\ref{BEL_life_insurance}). The distribution of change in BOFs is shown in Figure \ref{fig:changeBOF} where we observe a 99.5\% quantile, i.e.~the SCR, lying slightly above one million. If we did not consider parameter risk in the form of MCMC samples, the SCR would decrease by roughly 33\%.

\section{Model Validation}\label{validation}
 
Having estimated model parameters, it is straight-forward to derive validation techniques for our annuity model.
For the first procedure, transform the data by (\ref{transform})
such that we may (very roughly) assume that sequence of number of deaths $(N'_{a,g,k}(t))_{t\in\{1,\dots,T\}}$ is i.i.d.~over time. We then get explicit formulas for 
$\vars{N'_{a,g,k}}$, as well as $\cov(N'_{a,g,k}, N'_{a',g',k})$ and can then 
use samples from the Markov chains to derive quantiles. Then, these bounds can be compared to corresponding sample variances and sample covariances. 
In our Australian example 45.9 percent of all sample variances and
covariances lie within five and 95 percent quantiles.
	
For the second procedure define
\[
	N^*_{a,g,k}(t)
	:=\frac{N'_{a,g,k}(t)-\cexpvs{N'_{a,g,k}(t)}{\Lambda_k(t)}}
	{\sqrt{ \cvarsmall{N'_{a,g,k}(t)}{\Lambda_k(t)}}}
	=\frac{N'_{a,g,k}(t)-m_{a,g}\4q_{a,g}\4 w_{a,g,k}\4\Lambda_k(t)}
	{\sqrt{ m_{a,g}\4q_{a,g}\4 w_{a,g,k}\4\Lambda_k(t)}}\,,
\]
and note that the conditional 
central limit theorem implies $N^*_{a,g,k}(t)\to N(0,1)$ in distribution as $m_{a,g}(t)\to\infty$
where $N(0,1)$ denotes the standard normal distribution.
Knowing that $\cov(N^*_{a,g,k}(t), N^*_{a',g',k'}(t))=0$ for all $k\neq k'$, we 
may as well test for correlation with a simple $t$-test. Applying this procedure 
to Australian data, we get that 88.9 percent of all tests are accepted at a five percent significance level.

A third possibility is to test for serial correlation in 
$(N^*_{a,g,k}(t))_{t\in\{1,\dots,T\}}$, e.g., via the Breusch--Godfrey test. 
Applying this validation procedure on Australian data gives 
thatthe null hypothesis, i.e., that there is no serial correlation of order $1,2,\dots, 10$, is not rejected at a five percent level 
	 in 93.8 percent of all cases.
Serial correlation is interesting insofar as there may be serial causalities between a reduction in deaths due to certain death causes and a possibly lagged increase 
in different ones, see Figure \ref{fig:death_causes}. Note that we already remove a lot of dependence via time-dependent weights and 
death probabilities. 

Finally, we may use estimates for risk factor realisations to test whether 
risk factors are gamma distributed with mean one and variance $\sigma^2_k$ or not, 
e.g., via the Kolmogorov--Smirnov test. Note that estimates for risk factor realisations 
can either be obtained via MCMC based on the maximum a posteriori setting or by 
Equations (\ref{lambda_MAP}) or (\ref{MAPappr_lambda}). 
	For Australia, this test gives acceptance of
	the null hypothesis for all risk factors on all suitable levels of significance. 
	
	For choosing a suitable family for mortality trends, information criteria such as AIC, BIC, or DIC can be applied straight away. 
	 The decision how many risk factors to use cannot be answered by traditional
	information criteria since a reduction in 
risk factors leads to a different data structure.
It also depends on the ultimate goal. For example, if the development of all death causes is of interest, then a reduction of risk factors is 
not wanted. On the contrary, in the context of annuity portfolios several risk factors may be merged to one risk 
factor as their contributions to the risk of the total portfolio are small.

\section{Conclusion}\label{sec:conclusion}
Our approach provides a useful actuarial tool with numerous applications such as 
stochastic mortality modelling, P\&L derivation in annuity
and life insurance portfolios as well as (partial) internal model applications. Yet, there exists a fast and numerically
stable algorithm to derive loss distributions exactly, even for large portfolios.
We provide various estimation procedures based on publicly available data.
The model allows
for various other applications, including mortality forecasts. Compared to the
Lee--Carter model, we have a more flexible framework, get tighter bounds
and can directly extract several sources of uncertainty.
Straightforward model validation techniques are available.

\bibliography{pension}
\bibliographystyle{amsplain} 
\end{document}